\documentstyle[aps,prl,floats,epsfig]{revtex}

 \newcommand{\be}{\begin{equation}}
 \newcommand{\ee}{\end{equation}}
 \newcommand{\bea}{\begin{eqnarray}}
 \newcommand{\eea}{\end{eqnarray}}

 \newcommand{\ket}[1]{ | \, #1  \rangle}
 \newcommand{\bra}[1]{ \langle #1 \,  |}
 \newcommand{\proj}[1]{\ket{#1}\bra{#1}}

\begin{document}
  \begin{center}
                 {\bf  On the entanglement structure in
                 quantum  cloning  }\newline
  
                 {Dagmar~Bru\ss $^1$ 
                  and Chiara~Macchiavello$^2$}\newline
                 {\em
                 $^1$Inst. f\"{u}r Theoret. Physik, Universit\"{a}t Hannover,
                 Appelstr. 2, D-30167 Hannover, Germany\\
                 $^2$Dipartimento di Fisica ``A. Volta'' and INFM-Unit\`a 
                 di Pavia,
                 Via Bassi 6, 27100 Pavia, Italy}
  \end{center}
                 \date{Received \today}
                  \begin{abstract}

We study the entanglement properties of the output state of a universal 
cloning machine. We analyse in particular bipartite and tripartite
entanglement of the clones, and discuss the ``classical limit'' of
infinitely many output copies.

                 \end{abstract}
                PACS {03.67.-a, 03.65.Ud, 03.67.Mn} \newline
                 
\section{Introduction}

A fundamental law in quantum physics is the no-cloning theorem
\cite{nocwootters,yuen}, which tells us that it is impossible to
copy an unknown quantum state perfectly. This feature
is a direct consequence of the linearity of the Schr\"odinger equation.

From a physical point of view, 
it should better not be allowed to copy an unknown quantum state
perfectly, because otherwise a conflict with other 
well-established laws
of  physics would arise:
first, perfect quantum cloning would allow to use an entangled
EPR-state that is shared between two remote
parties to send information across the distance of the parties,
where the speed of the signalling would only depend on the speed
of the cloning machine and a subsequent measurement,
thus allowing superluminal signalling.
Second, perfect cloning would allow to produce infinitely many
identical copies of a given spin state, half of which could be measured
e.g. in the $x$-basis, half of them in the $y$-basis, therefore
allowing to measure the expectation
value of non-commuting observables simultaneously.

Since the beginning of the development of the field of quantum 
information 
the no-cloning theorem has also been viewed as one of the most
fundamental differences between classical and quantum information 
theory. The impossibility of perfectly cloning an unknown quantum state
has far-reaching consequences: for instance, it disables a spy from copying a
quantum signal and resending the original, and is thus the reason
for the security of  quantum cryptography. At the same time, however,
it also disables us from making a ``quantum back-up'', i.e. to
keep a copy of an unknown quantum state for the purpose of
error correction.

The no-cloning theorem only tells us that it is impossible make
a {\em perfect} copy of an unknown quantum state. In the recent
years, much work has been done in order to explore the limits
for {\em approximate} cloning transformations
\cite{buzek-hill,gima,oxibm,bc,werner,prob}. 
This is often loosely referred to as quantum cloning.
Experimental realisations of quantum cloning have been suggested
in quantum optics 
\cite{simon} and cavity QED \cite{raimond},
and experiments have been performed in quantum optics
\cite{dik} and nuclear magnetic resonance experiments
\cite{jones}.

Here, we  want to 
investigate an aspect of 
quantum cloning that has not received any attention so far,
namely we want to 
study the entanglement properties of the output state of a cloning
device.
The motivation for this study is two-fold: first, we wish to 
understand some fundamental properties of quantum cloning, by
shedding
light on a certain limit of quantum cloning -- the 
case of producing infinitely many clones. 
It was shown that the optimal fidelity of
  this case corresponds to the optimal
state estimation process \cite{bc}, i.e. a cloner 
with infinitely many copies could be implemented by 
optimally estimating the input
and then producing a product state of infinitely many copies
by using this knowledge. In this sense this case is referred to
as ``classical limit'', see also \cite{gima}.
Here we want to ask the
question: is it justified to call the case of infinitely many copies
the classical limit,
when judging by the entanglement structure of the
output  state? 

Second, we want to
investigate the multipartite entanglement properties of the
cloning output. One  arrives at the shape of this
 specific entangled state by
maximising a scalar quantity, namely the fidelity, i.e. the overlap
$F=\bra{\psi}\varrho_{out}\ket{\psi}$, where $\ket{\psi}$ is
the input state and $\varrho_{out}$ the one-particle reduced
density operator of the output.
Does the cloning output have some significance in nature, i.e. does
it resemble
a state that appears naturally, e.g. in the ground state of
some quantum statistical system? This question is motivated by recent
studies of multiparticle entanglement 
in this direction \cite{chain,nielsen,vedral,fazio}.

\section{Entanglement properties of  the $1\to 2$ cloning state}

Let us commence by looking into the most simple case, namely the
output of a universal cloner that takes any pure state of
one qubit as input and produces
two identical copies with optimal fidelity 5/6. 
In the next section  we will then look at general universal cloning
transformations. Throughout this paper we will only consider
two-dimensional states and 
symmetric universal cloning transformations, i.e. 
transformations 
for which the fidelity of all output copies is equal and their
 quality does not depend on the form of the input state. 
For these universal transformations we can study
without loss of generality
the entanglement structure of the cloning output for an input
that is the basis state $\ket{0}$. This is given by
\be
{\cal U}\ket{0}\ket{0}\ket{a}=\sqrt{\frac{2}{3}}\ket{0}\ket{0}\ket{0}
+\sqrt{\frac{1}{6}}(\ket{0}\ket{1}+ \ket{1}\ket{0})\ket{1}\ ,
\label{3pure}
\ee
 where the first qubit is the original, the second is
the clone and the third  is an auxiliary system, usually called ancilla. 
Clearly, the total state is entangled. How is the entanglement
distributed over the state, i.e. how much are the two clones
or a clone and an ancilla entangled with each other?
The reduced density matrices  $\varrho_{cc}$ for  two clones 
and $\varrho_{ca}$  for one clone
and one ancilla are given by
\be
\varrho_{cc} = \left(
\begin{array}{cccc}\frac{2}{3} & 0 & 0 & 0 \\
    0  & \frac{1}{6} & \frac{1}{6} & 0 \\
     0 & \frac{1}{6} & \frac{1}{6} & 0 \\
      0 & 0 & 0 & 0 \\
\end{array}\right) \ \ \ \  ;  \ \ \ \
\varrho_{ca} = \left(
\begin{array}{cccc}\frac{2}{3} & 0 & 0 & \frac{1}{3} \\
    0  & \frac{1}{6} & 0 & 0 \\
     0 & 0 & 0 & 0 \\
       \frac{1}{3} & 0 & 0 &  \frac{1}{6} \\
\end{array}\right)\ .
\label{denscc}
\ee

Thanks to the analytical
 formula for the concurrence by Wootters \cite{wootters}
it is easy to calculate the entanglement of formation for these
bipartite density matrices. As shown in \cite{wootterser},
a density matrix of the shape
\be
\sigma = \left(
\begin{array}{cccc}a & 0 & 0 & 0 \\
    0  & b & c & 0 \\
     0 & c^* & d & 0 \\
      0 & 0 & 0 & e \\
\end{array}\right)
\label{density}
\ee
in the basis $\{\ket{00},\ket{01},\ket{10},\ket{11}\}$
 has the concurrence 
\be
C(\sigma)= 2\,  max\,  (|c|-\sqrt{ae},0) \ .
\label{conc}
\ee
The entanglement of formation $E_F$ can then be expressed as
a function of $C$, namely
\be
E_F=-\frac{1+\sqrt{1-C^2}}{2}\log_2 \frac{1+\sqrt{1-C^2}}{2}
-\frac{1-\sqrt{1-C^2}}{2}\log_2 \frac{1-\sqrt{1-C^2}}{2} \;.
\label{ef}
\ee
In the  case of the cloning output from equation (\ref{denscc})
this leads to $C_{12}=1/3$ for the two clones and
$C_{23}=2/3$ for clone and ancilla. Using the relation (\ref{ef}) between
the concurrence and the entanglement of formation  we
arrive at $E_{F,12}\simeq 0.1873$ for the state of the two clones and
$E_{F,23}\simeq 0.55$ for the state of one clone and the ancilla. 
Note that the entanglement 
between clone and ancilla is higher than between the two clones.

Let us also study the entanglement properties of the total pure
state (\ref{3pure}). It is obviously genuinely three-party entangled, but there
exist two different inequivalent classes of three-party entangled
states \cite{W}, namely 
the W- and the GHZ-class -- to which one does it
belong? 
Let us remind the reader  that states of the form
\be
\ket{GHZ}=\frac{1}{\sqrt 2}(\ket{000}+\ket{111})
\ee
are called GHZ states (these states can be detected by the Mermin 
inequality \cite{mermin}),  
and states of the form 
\be
\ket{W}=\frac{1}{\sqrt 3}(\ket{001}+\ket{010}+\ket{100})
\ee
are called W states \cite{W}.
For pure states of three qubits there is a simple criterion to establish
whether an entangled state belongs to the GHZ class or to the W class,
namely the 3-tangle \cite{3tangle}. It is straightforward to verify that the 
3-tangle is zero for
the state (\ref{3pure}), and therefore we can conclude that 
the total cloning output
is  in the
W-class. We want to point out that this is very reasonable: 
the 1-particle reduced density matrices
for GHZ-states are maximally mixed, and therefore could not
represent high fidelity copies.

Finally, we point out that the state (\ref{3pure}) happens to have the 
property that the three tangles  add up to 1, i.e.
$C_{12}^2+C_{13}^2+C_{23}^2=1$.
This value lies  the interval for this sum that
is allowed by the sum rules of \cite{3tangle}. Remember that for
a GHZ state the tangles add up to zero, because the bipartite
reduced density matrices are separable, and they add up to 4/3
for the W state.

\section{Bipartite entanglement in the output of a  universal cloner}
\label{bipartite}

Let us now generalise the above ideas to an $N\to M$-cloner
for qubits, i.e. a cloner that takes $N$ inputs and produces
$M$ outputs. What is our expectation for this case? How
will the bipartite entanglement scale with the number $M$ of
outputs? The case $M\to \infty$ is often referred to as classical
limit, and it has indeed been shown in  \cite{bc} that the fidelity
for producing infinitely many copies can be also reached by
making an optimal state estimation on the input and then producing $M$
identical uncorrelated states that correspond to the guessed state.

The optimal cloning transformation is given by \cite{gima}
\be
U_{N,M}\ket{N\psi}\ket{(N-M)s}\ket{R}=
\sum_{j=0}^{M-N}\alpha_j(N,M)\ket{(M-j)\psi,j\psi^\perp}\ket{R_j(\psi)}\;,
\label{nmcloner}
\ee
where $\ket{N\psi}$ is the state of $N$ qubits all in state $\ket{\psi}$,
$\ket{s}$ and $\ket{R}$ are arbitrary input states of the $M-N$ copies and
the state of the ancilla respectively, 
$\ket{k\psi,j\psi^\perp}$ is the normalised symmetric state
of $k+j$ qubits (with $k$ qubits 
in state $\ket{\psi}$ and $j$ qubits in the orthogonal state 
$\ket{\psi^\perp}$), and 
\be
\alpha_j(N,M)=\sqrt{\frac{N+1}{M+1}}\sqrt{\frac{(M-N)!(M-j)!}{(M-N-j)!M!}}\;.
\label{alpha}
\ee
The output states of the ancilla are composed of $M-1$ qubits and
are given by 
\be
\ket{R_j(\psi)}=\ket{(M-1-j)\psi^*,j(\psi^*)^\perp}\;,
\label{ancilla}
\ee
where $\psi^*$ denotes the complex conjugation of the wavefunction $\psi$.

This transformation is symmetric in the space of the clones,
and therefore leads to the same fidelity for all clones. The
symmetry of this transformation also means that the reduced 
density matrix for two clones will be of the form (\ref{density}),
with the further simplification of $b=c=d$.
Note that {\em any} two clones will have the same reduced density
matrix.

From (\ref{nmcloner}) one calculates the elements of the 
reduced bipartite density matrix, with the notation as
in (\ref{density}):
\bea
a(N,M)&=&
 \sum_{j=0}^{M-N}\alpha_j^2(N,M) \cdot \frac{(M-j)(M-j-1)}
 {M(M-1)}\ , \nonumber \\
c(N,M)&=& 
\sum_{j=0}^{M-N}\alpha_j^2(N,M) \cdot \frac{j(M-j)}
 {M(M-1)}\ , \nonumber \\
e(N,M)&=& 
\sum_{j=0}^{M-N}\alpha_j^2(N,M) \cdot \frac{j(j-1)}
 {M(M-1)} \ .
\label{aec}
\eea
What do these coefficients tell us? Let us first study in detail the 
simple case of $N=1$, extending the results of the previous section to an 
arbitrary number of output copies. For
$N=1$ the above coefficients take the form
\bea
a(1,M)= \frac{3M+2} {6M}\;,\qquad
c(1,M)= \frac{1}{6}\;,\qquad
e(1,M)= \frac{M-2}{6M}\;.
\eea
As we can see, the concurrence surprisingly vanishes for $M\geq 3$.
In the case of universal cloning with $N=2$ we have a similar behaviour:
the coefficients take the form
\bea
a(2,M)= \frac{3M^2-2} {5M(M-1)} \;,\qquad
c(2,M)= \frac{3M^2-5M-2}{20M(M-1)} \;,\qquad
e(2,M)=\frac{M^2-5M+6}{10M(M-1)}\;.
\eea
As we can easily verify, the concurrence vanishes for $M\geq 4$.

For a generic value of $N$ it is not straightforward to discuss the form of the
coefficients (\ref{aec}). However, we can analyse the limit $M\to\infty$.
Each of the coefficients consists
of a finite sum, let us call the terms in the sum
$a_j,c_j$ and $e_j$. How does $\sqrt{ae}$ compare with
$c$? For large $M$  we find that  $\sqrt{a_je_j}  \approx c_j$,
and therefore we find for the concurrence
\be
C= 2\, max\, (0-\sqrt{\sum_{i\neq j}a_ie_j},0)=0 \ .
\label{concinfty}
\ee
This means that our intuition about any two clones being in a 
separable state for $M\to \infty$ is confirmed. However, we also
realise that in this limiting case  $\sqrt{ae}$ is much larger
than $c$. Maybe the concurrence already vanishes for a finite $M$,
and if so, for which? 
We have seen that for $N=1$ the concurrence vanishes already for $M=3$, and
for $N=2$ it vanishes for $M=4$. So, does this hold in general,
i.e. is there  no entanglement for the  $N\to N+2$-cloner?
Well, first of all we can easily see that in the $N\to N+1$ case there is 
always entanglement, because $e(N,N+1)=0$ and $c(N,N+1)\neq 0$.
Moreover, the explicit calculation of the $N\to N+2$ case gives the following 
results
\bea
a(M-2,M)= \frac{M^4-5M^2+8} {M^2(M^2-1)} \;,\qquad
c(M-2,M)= \frac{2(M^2-3)}{M^2(M^2-1)} \;,\qquad
e(M-2,M)= \frac{4}{M^2(M^2-1)}\;.
\eea
As we can see, the corresponding concurrence is always zero for $M\geq 3$,
and it is therefore reasonable to expect that entanglement really vanishes
when the number of output copies exceeds the number of inputs by 
at least two.


One can  also study the entanglement between a clone
and the ancilla. 
Here we use the form of the ancilla as given in equation (\ref{ancilla}).
The density matrix of an output copy and an ancilla qubit can be also written
in the form (\ref{density}), where now the basis
is  $\{\ket{01},\ket{00},\ket{11},\ket{10}\}$:
\bea
a(N,M)&=& f(N,M)
\sum_{j=0}^{M-N}j(M-j)\frac{(M-j)!}{(M-N-j)!}\ ,
\nonumber \\
b(N,M)&=& f(N,M)
\sum_{j=0}^{M-N}(M-j)(M-j-1)\frac{(M-j)!}{(M-N-j)!}\ ,
\nonumber \\
c(N,M)&=& f(N,M)
\sum_{j=0}^{M-N-1}(j+1)\sqrt{\frac{M-j-1}{M-N-j}}\frac{(M-j)!}{(M-N-j-1)!}\ ,
\nonumber \\
d(N,M)&=& f(N,M)
\sum_{j=0}^{M-N}j^2\frac{(M-j)!}{(M-N-j)!}\ ,
\nonumber \\
e(N,M)&=& f(N,M)
\sum_{j=0}^{M-N}j(M-j-1)\frac{(M-j)!}{(M-N-j)!}\ .
\label{matanc}
\eea
where $f(N,M)=\frac{N+1}{M(M^2-1)}\frac{(M-N)!}{M!}$. 
Again, for generic $N$ we can only study the limit $M\to \infty$,
using a similar argument as for the entanglement between the clones:
we label the $j$th term in the sum for the coefficient $a$ in equation
(\ref{matanc}) $a_j$, and accordingly for coefficients $c$ and $e$.
Then for infinitely large $M$ we find that $c_j^2\leq a_je_j$, and
therefore the concurrence vanishes.

\begin{figure}[ht]
\begin{picture}(150,150)
\put(8,5){\epsfxsize=380pt\epsffile[23 146 546 590 ]{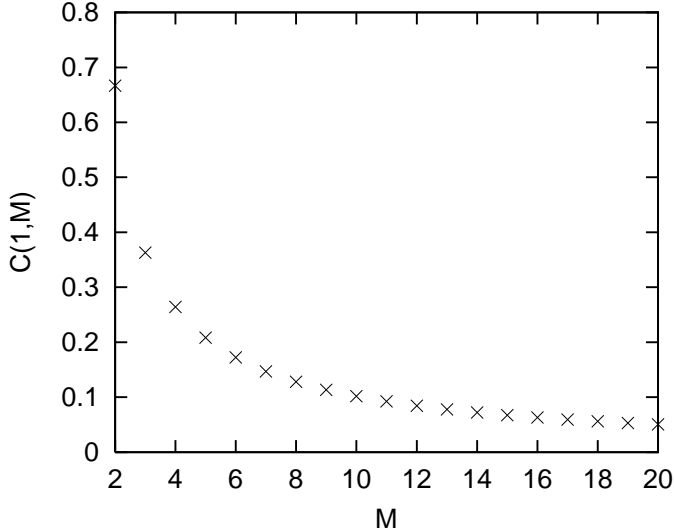}}
\end{picture}
\vspace*{3.cm}
\caption[]{\small 
Concurrence $C(1,M)$ for the state of a clone and an ancilla
qubit as a function of the number of output copies $M$. The corresponding
analytic form is given in Eq. (\ref{ancn=1}).}
\label{f:fig1}
\end{figure}

The form of the coefficients in this case is more complicated than
before. As an illustration, we report the analytic results
for the case $N=1$, where the above coefficients take the simpler form
\bea
a(1,M)&=& d(1,M)=\frac{1}{6}\;,\qquad
b(1,M)= \frac{3M+2}{6M}\;,\nonumber\\
c(1,M)&=& \frac{M+2}{6M}\;,\qquad
e(1,M)= \frac{M-2}{6M}\;.
\label{ancn=1}
\eea
The concurrence in this case is given by
\bea
C(1,M)=\frac{1}{3}\left(\frac{M+2}{M}-\sqrt{\frac{M-2}{M}}
\right)\;.
\label{anconc}
\eea
As we can see, $C(1,M)$ is always different from zero for any value of $M$  
and vanishes in the limit $M\to\infty$. The behaviour of $C(1,M)$ is plotted in
Fig. \ref{f:fig1}. This result is a bit surprising: while correlations
of quantum nature 
vanish very quickly with increasing $M$ for
 two output copies, they survive 
for all finite values of $M$ for states of one output clone
and one ancilla.

Let us interpret and summarise our results. 
The global output state of a universal
symmetric quantum cloner is an entangled state. However, the clones
in general do not carry any bipartite entanglement, unless we consider
a very low number of copies: we have found that the entanglement does
 $\em not$ gradually decrease with the number of copies, but vanishes
already 
for $M=3$, if one input is given, and for $M=4$, if two inputs are given.
Therefore, judging only by the bipartite entanglement properties, one 
{\em cannot}
conclude that the limit $M\to \infty$ corresponds to the ``classical''
limit, as the quantum part of the bipartite 
correlations already vanishes for a
very low number of outputs. 
We also found that the 
cloning states do not resemble the ground states of quantum
statistical systems, which are typically either separable states or
bear bipartite entanglement, whereas here a generic state is
entangled, but without bipartite entanglement between the clones.

In order to characterise the entanglement structure of the
global cloning state fully, one would need to study the entanglement
properties of subsystems of any size. In the following section
we consider the case of a tripartite subsystem. 

\section{Tripartite entanglement in the output of a universal cloner}

 In section \ref{bipartite}
we have shown that the total output of a $1\to 2$ cloner,
that is a pure state of two clones and the ancilla, corresponds
to a W-state. If there is no bi-partite entanglement for the
clones in a $1\to 3$ cloner, maybe there is tripartite entanglement?
Unfortunately,
the problem of characterising the entanglement properties is not
solved for general mixed states of three qubits. However, we
will show some partial results for the cloning states.

Let us study the reduced density matrix $\varrho_{123}(N,M)$
for three clones in
general, i.e. as a function of the number $N$ of inputs and
number $M$ of outputs. From equation (\ref{nmcloner}) we find
\bea
\varrho_{123}(N,M) &=&
 \sum_{j=0}^{min[M-N,M-3]}\alpha_j^2 \cdot
\frac{{M-3 \choose j}}{{M\choose j}}
 \proj{000}+
 \sum_{j=1}^{min[M-N,M-2]}\alpha_j^2 \cdot
\frac{{M-3 \choose j-1}}{{M\choose j}} \cdot
3 P_{W_{100}}+\nonumber \\
&& \sum_{j=2}^{min[M-N,M-1]}\alpha_j^2 \cdot
\frac{{M-3 \choose j-2}}{{M\choose j}} \cdot
3 P_{W_{110}}+
 \sum_{j=3}^{M-N}\alpha_j^2 \cdot
\frac{{M-3 \choose j-3}}{{M\choose j}}
 \proj{111}\ ,
\label{three}
\eea
where $\alpha_j$ depends on $N$ and $M$ and was defined
 in equation (\ref{alpha}),
and $P_{W_{100}}$ denotes the projector onto the pure W state
where one of the three qubits is in the state 1.
 In general, it is not known for which 
coefficients a mixture of this type is entangled. Therefore,
let us restrict to two special cases.

The reduced density matrix for the three clones in the $1\to 3$
cloner is found to be
\be
\varrho_{123}(1,3) = \frac{1}{2}\proj{000}+\frac{1}{3} P_{W_{100}}
+\frac{1}{6} P_{W_{110}}\ .
\label{tripartite}
\ee
This is a mixture  of two different W-states and a product state.
Therefore, according to the classification of \cite{Acin} it is
at most in the W class, i.e. not in the GHZ class.

When one wants to detect
the entanglement properties of multipartite states, the Mermin
inequalities immediately \cite{mermin} come to ones mind. 
However, in their original form they are violated by the GHZ states,
but satisfied by  the W states: these two inequivalent classes of
states show different forms of the violation of local realism.
A generalisation of the Mermin
inequality which is also  violated for W states was introduced
in \cite{cabello}. However, testing this generalised
Mermin inequality for the
state from equation (\ref{tripartite}) does not show a  violation.

Let us also study the case of  the $2\to 3$
cloner, where 
the reduced density matrix for the three clones  is given by
\be
\varrho_{123}(2,3) = \frac{3}{4}\proj{000}+\frac{1}{4} P_{W_{100}}
\ ,
\label{tripartite2}
\ee
which is a mixture  of only one  W state and a product state.
Again, it  is
at most in the W-class, i.e. not in the GHZ class. And again, 
the generalised Mermin inequality from  \cite{cabello}
is not violated by this state.

As we did not find a violation of a Mermin inequality, the 
three-qubit states given above could be in principle separable,
bi-separable or in the W-class. However, we can also study the
partial transpose  with respect to Alice
of a state of the general form
\be
\varrho_{123}(N,M)=
p_0
 \proj{000}+p_1
 P_{W_{100}}+
p_2
 P_{W_{110}}+
p_3
 \proj{111}\ ,
\label{pi}
\ee
with $p_i>0$ and $\sum_i p_i=1$.
 Notice that 
 due to symmetry all three partial transposes
with respect to one subsystem have the same structure.
 In general,
the partial transpose turns out to be  non-positive iff
either $p_1^2>3p_0p_2$ or $p_2^2>3p_1p_3$.
This means  that for either $p_0=0$ or $p_3=0$ the 
 partial
transpose
is non-positive, independently of the other probabilities.
 Therefore the cloning   states given in equations (\ref{tripartite})
and (\ref{tripartite2}) cannot be
separable \cite{Asher}. Indeed, although there is no bipartite
entanglement in the $1\to 3$-cloner, we have found that 
there is tripartite
entanglement. We also see immediatley from equation (\ref{three})
that for the $N\to N+1$ cloner 
the partial transpose is always non-positive,
as the last two terms in the equation
vanish. Furthermore, for the $N\to N+2$ cloner 
 the last term in equation (\ref{three})
vanishes, and thus there is  always tripartite 
entanglement for this case, contrary to the bipartite case.

Let us finally study tripartite entanglement for the general case of one
input, i.e. the $1\to M$-cloner, especially 
in the limit $M\to \infty$. The corresponding weights $p_i$, which
were defined in equation (\ref{pi}), are given by
\bea
p_0(1,M)= \frac{4M+3}{10M} \;,\qquad
p_1(1,M)= \frac{3M+1}{10M} \;,\qquad
p_2(1,M)= \frac{2M-1}{10M} \;,\qquad
p_3(1,M)= \frac{M-3}{10M}  \;.
\eea
For $M\to \infty$ one finds immediately that the partial transposes
are positive, and thus there is no free tripartite entanglement
for infinitely many copies. In fact,  for $M>4$ the partial tranposes
are positive, and thus tripartite entanglement vanishes already
for small numbers of output copies, analogously to the bipartite case.

\section{Conclusions}

In summary, we have studied the entanglement structure at the
output of an approximate cloning device. Regarding bipartite 
entanglement between two cloning outputs, we have found the
surprising result that it vanishes
 for the $N\to N+2$ cloner.
Thus,
bipartite entanglement does not only disappear for 
the ``classical'' limit of infinitely many copies, but already 
when 
the number of outputs exceeds the number of inputs by two. 
Using one input, we showed that
the entanglement between clone and ancilla, however, only vanishes
in the limit of infinitely many copies.
We also proved on the other hand that tripartite entanglement is always
present for the $N\to N+2$ cloner. This entanglement 
generically seems to be
of the W type, rather than of the GHZ type.
We hope that these results will be useful for the further
understanding of multipartite entanglement.

This work was supported in part by the European Union projects EQUIP
(contract IST-1999-11053) and ATESIT (contract IST-2000-29681).
DB acknowledges support from the ESF Programme QIT
and DFG-Schwerpunkt QIV.


\begin{thebibliography}{99}

\bibitem{nocwootters} W. Wootters and W. Zurek, Nature {\bf 299}, 802 (1982).
\bibitem{yuen} H.P. Yuen, Phys. Lett. A {\bf 113}, 405 (1986).
\bibitem{buzek-hill} V.~Bu\v{z}ek and M.~Hillery, Phys. Rev. A {\bf 54}, 
1844 (1996).
\bibitem{gima} N.~Gisin and S.~Massar, Phys.~Rev.~Lett. {\bf 79}, 2153
                   (1997).
\bibitem{oxibm} D.~Bru\ss, D.~DiVincenzo, A.~Ekert, C.~Fuchs,
             C.~Macchiavello
               and J.~Smolin, Phys.~Rev. A {\bf 57}, 2368 (1998).
\bibitem{bc} D.~Bru\ss , A.~Ekert, C.~Macchiavello, 
         Phys. Rev. Lett.  {\bf 81}, 2598 (1998).
\bibitem{werner} R. Werner, Phys.~Rev. A {\bf 58}, 1827 (1998).
\bibitem{prob} L.-M. Duan and G.-C. Guo,
 Phys. Rev. Lett. {\bf  80}, 4999 (1998).
\bibitem{simon}  C. Simon, G. Weihs, and A. Zeilinger,
             Phys. Rev. Lett. {\bf 84}, 2993 (2000).
\bibitem{raimond}  P. Milman, H. Ollivier, J. M. Raimond, quant-ph/0207039.
\bibitem{dik}  A. Lamas-Linares, C. Simon, J. C. Howell, and D. Bouwmeester,
 Science {\bf 296}, 712 (2002).
\bibitem{jones}   H. K. Cummins, C. Jones, A. Furze, N. F. Soffe,
 M. Mosca, J. M. Peach, and J. A. Jones, 
             Phys. Rev. Lett. {\bf 88}, 187901 (2002).
\bibitem{chain}  K. M. O'Connor and W. K. Wootters,
 Phys. Rev. A {\bf 63}, 052302 (2001).
\bibitem{nielsen}  T. J. Osborne and M. A. Nielsen, quant-ph/0202162.
\bibitem{vedral}  M. C. Arnesen, S. Bose, and V. Vedral, 
             Phys. Rev. Lett. {\bf 87}, 017901 (2001).
\bibitem{fazio} A. Osterloh, L. Amico, G. Falci, and  R. Fazio,
 Nature {\bf 416}, 608 (2002).
\bibitem{wootters} W.K. Wootters,
 Phys. Rev. Lett.{\bf  80}, 2245 (1998).
\bibitem{wootterser} K.M. O'Connor and W.K. Wootters,
 Phys. Rev. A {\bf  63}, 052302 (2001).
\bibitem{W} W. D\"ur, G. Vidal and  J.I. Cirac
 Phys. Rev. A {\bf  62}, 062314 (2000).
\bibitem{mermin} N. D. Mermin, Phys.~Rev. Lett. {\bf 65}, 1838 (1990).
\bibitem{3tangle} V. Coffman, J. Kundu, and W. Wootters,
 Phys. Rev. A {\bf  61}, 052306 (2000).
\bibitem{Acin} A. Acin, D.~Bru\ss, M. Lewenstein and A. Sanpera.
Phys. Rev. Lett. {\bf  87}, 040401 (2001).
\bibitem{cabello} A. Cabello, Phys. Rev. A {\bf 65}, 032108 (2002).
\bibitem{Asher} A. Peres, Phys. Rev. Lett. {\bf  77}, 1413 (1996).
\end{thebibliography}
\end{document}